\documentclass[pre,twocolumn]{revtex4}
\usepackage{graphicx}
\usepackage{epstopdf}
\usepackage{amsmath}
\usepackage{titlesec}
\usepackage{tikz}
\usepackage{array}
\usetikzlibrary{shapes,arrows,external}
\newcommand{\bR}{{\bf R}}

\begin{document}
	
%\title{Using Fluctuations of the Local Energy to Improve Many-Body Wave Functions}

%\title{Using Correlated Local Operator Fluctuations to Diagnose Wave Function Deficiencies}

\title{Using Local Operator Fluctuations to Identify Wave Function Improvements}

\author{Kiel T. Williams}
\author{Lucas K. Wagner}
\email{lkwagner@illinois.edu}
\affiliation{Department of Physics; University of Illinois at Urbana-Champaign, Urbana, IL}

\begin{abstract}
A method is developed that allows analysis of quantum Monte Carlo simulations to identify errors in trial wave functions. The purpose of this method is to allow for the systematic improvement of variational wave functions by identifying degrees of freedom that are not well-described by an initial trial state. We provide proof of concept implementations of this method by identifying the need for a Jastrow correlation factor, and implementing a selected multi-determinant wave function algorithm for small dimers that systematically decreases the variational energy. Selection of the two-particle excitations is done using quantum Monte Carlo within the presence of a Jastrow correlation factor, and without the need to explicitly construct the determinants. We also show how this technique can be used to design compact wave functions for transition metal systems. This method may provide a route to analyze and systematically improve descriptions of complex quantum systems in a scalable way.
\end{abstract}

\maketitle

\section{Introduction}

First principles quantum Monte Carlo calculations\cite{foulkes_quantum_2001} for solids are a promising way to go beyond density functional theory (DFT).
These methods directly simulate electron-electron correlations and can obtain very high accuracy on challenging materials\cite{wagner_quantum_2014,wagner_ground_2015,foyevtsova_textitab_2014,zheng_computation_2015} using current state of the art techniques like fixed node diffusion Monte Carlo (DMC).
Despite this success, the DMC method's accuracy is limited by the fixed node approximation, which allows for polynomial scaling of the computational cost with system size, but results in a DMC energy that is only an upper bound to the true ground state energy.
In practical calculations, improvement of the accuracy and efficiency of fixed node diffusion Monte Carlo is reliant on improving trial wave functions which determine the fixed nodal surface.

In order for a trial wave function to be appropriate for quantum Monte Carlo calculations, it should be compact and efficient to calculate. 
For application to bulk materials, it must also be size-extensive; that is, the total energy must scale with the system size. 
By far the most common trial wave function is the Slater-Jastrow wave function\cite{jastrow_many-body_1955,ceperley_monte_1977}, which is simple, extensive, and initial guesses are easily obtainable from DFT codes.
While truncated determinant expansions can be effective in describing small molecules\cite{petruzielo_approaching_2012}, they cannot be used in bulk materials because they are not size-extensive.
Backflow wave functions\cite{feynman_energy_1956} have proven effective in homogeneous\cite{kwon_effects_1998} and inhomogeneous\cite{lopez_rios_inhomogeneous_2006} systems, but worsen the computational scaling with system size of QMC methods and sometimes do not offer improvement of results.
It is thus of great interest, given a Slater-Jastrow wave function, whether there is a compact wave function that describes the most important improvements relative to the ground state.

In this article, we present some first steps towards a method that uses fluctuations of the local energy $\hat{H}\Psi(\bR)/\Psi(\bR)$, not to optimize a given parameterization, but to identify directions in Hilbert space that can improve trial wave functions.
We first provide a summary of the principle of using the imaginary time projector $\exp(-\tau \hat{H})$ to improve wave functions and the notation that will be used in the article.
Then we show a proof of concept for multi-Slater Jastrow wave functions, in which this method is used to select determinants in the wave function.
Finally, we show how the local energy fluctuations can be used to determine {\it a priori} what terms to add to a variational wave function for a transition metal system TiO.
These results set the stage for data mining of many-body wave functions to determine how they should be improved.

\section{Theory}

\begin{figure}
	\includegraphics{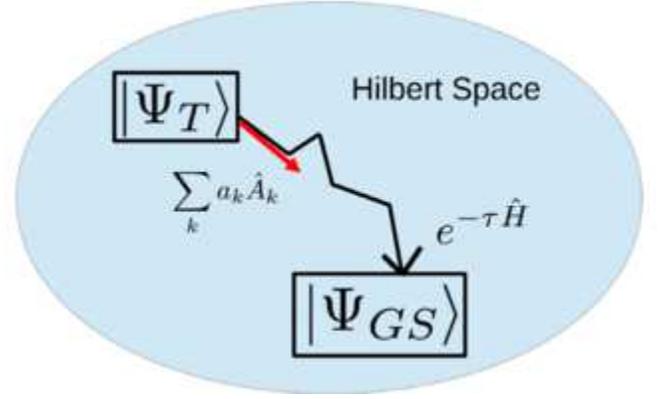}
  \caption{Visual representation of the path through Hilbert space from the initial to exact wave function taken by the exact (black) and mimicked (red) projection operators respectively.}
	\label{hilbert_cartoon}
\end{figure} 

In this work, we use ideas that have been known for a long time for optimizing parameters in wave functions\cite{neuscamman_optimizing_2012,toulouse_optimization_2007,filippi_optimal_2000}, but we follow more the work of Holtzmann\cite{holzmann_backflow_2003} in that we would like to use the Feynman-Kac formulae to discover which parameterization to add to a given wave function.
The quantum variational principle states that for any appropriately normalized trial wave function $\Psi_{T}(\textbf{R}, P)$, where $\textbf{R}$ is the many-body electron coordinate and $P$ is a set of parameter values, the expectation value of the Hamiltonian of the system in state $\Psi_{T}$ equals or exceeds the ground state energy of the Hamiltonian:
\begin{equation}
\label{eq:1}
E_{T} \geq E_{g},
\end{equation}
where:
\begin{equation}
\label{eq:2}
E_{T} (P) = \langle \Psi_{T} | \hat{H} | \Psi_{T} \rangle.
\end{equation}
We then minimize  $E_T(P)$ with respect to the parameter set $P$. 
Once this is done, we must alter the parameterization of the trial wave function to obtain further improvement in the energy estimate. 
Our ultimate goal will be not to optimize the parameters within a fixed set $P$, but to identify new parameters that must be added to $P$ to improve the qualitative structure of the particular trial state.

Iteratively applying the projection operator to a trial function $\Psi_{T}$ produces a sequence of new wave functions:
\begin{equation}
\label{eq:2.5}
| \Psi_{H}( \tau) \rangle = e^{- \tau \hat{H}} | \Psi_{T} \rangle .
\end{equation}
This converges to the exact ground-state wave function $|\Psi_{GS}\rangle$ in the infinite limit:
\begin{equation}
\label{eq:4}
\lim_{\tau\rightarrow\infty} | \Psi_{H} ( \tau ) \rangle = | \Psi_{GS} \rangle ,
\end{equation}
%The exact ground state can be obtained through an iterative process by taking a small-$\tau$ limit.  
%To first-order:
%\begin{equation}
%\label{eq:3}
%| \Psi_{H}(  \tau) \rangle = (1-  \tau \hat{H}) | \Psi_{T} %\rangle.
%\end{equation}
Performing this operation directly corresponds to a projector Monte Carlo method, such as diffusion Monte Carlo. 
The challenge in doing this is that compact representations of the operator $\exp(-\tau\hat{H})$ are generally not known, and so the imaginary time dynamics must operate in very high dimensions. Our objective here will be to find a compact representation of the short-time projector operator.

We begin by considering an arbitrary set of linear operators $\{ \hat{A}_{i} \}$. Iteratively applying this set of operators generates a new sequence of wave functions:
\begin{equation}
\label{eq:5}
| \Psi_{A} \rangle = \left ( 1 + \sum_{i} a_{i} \hat{A}_{i} \right ) | \Psi_{T} \rangle .
\end{equation}
For brevity, we define:
\begin{equation}
\label{eq:5_1}
| \Psi_{A_i} \rangle \equiv \hat{A}_i  | \Psi_{T} \rangle .
\end{equation}
We force the minimal set of operators $\{ \hat{A}_{i} \}$ to mimic the projection operator by minimizing the square deviation of $\Psi_{A}$  from $\Psi_{H}$:
\begin{equation}
\label{eq:6}
\int (\Psi_{A}(\textbf{R})-\Psi_{H}(\textbf{R}))^{2} d\textbf{R}.
\end{equation}
This minimization procedure provides an estimate of the set of associated $\{a_i\}$ operator amplitudes. We define the local operators $A_{k} (\textbf{R}) \equiv  \frac{\Psi_{A_k}(\textbf{R})}{\Psi_{T}(\textbf{R})}$ and a local energy $E_{L}(\textbf{R}) = \frac{\hat{H} \Psi (\textbf{R}) }{\Psi(\textbf{R})}$. By expanding the projection operator to first-order and minimizing the square deviation, we find that:
\begin{equation}
\label{7_5}
a_k = - \tau \int \frac{\hat{H} \Psi_{T}(\textbf{R})}{\Psi_{T}(\textbf{R})} \frac{\hat{A}_k \Psi(\textbf{R})}{\Psi_{T}(\textbf{R})} \Psi_{T}^{2}(\textbf{R}) d\textbf{R}.
\end{equation}
\begin{equation}
\label{7_6}
\frac{d a_k}{d \tau} = - \langle (E_L(\textbf{R})- \langle \hat{H} \rangle ) \hat{A}_{k} (\textbf{R}) \rangle,
\end{equation}
where we have assumed that elements of the set $\{ \Psi_{A_i} \}$ are orthogonal, or approximately orthogonal, such that $ \langle \Psi_{A_i} | \Psi_{A_k} \rangle \approx \delta_{ik} $. Fig. \ref{hilbert_cartoon} depicts this scheme pictorially, with the exact and mimicked projection operators represented by the black and tangential red arrows respectively. We see then that the mimicked projection operator evaluated for $\tau = 0$ can be viewed as a linearized approximation to the exact path to the ground state through Hilbert space. In this way, our approximation to the projection operator identifies the most significant elements of Hilbert space absent from an initial trial state. 

The derivation of our method is similar in spirit to the stochastic reconfiguration (SR) of Sorella \cite{sorella_green_1998,sorella_green_2000,sorella_generalized_2001,sorella_weak_2007,neuscamman_optimizing_2012}. The energy fluctutation potential method (EFP) also shares some similarities with our technique in its focus on the correlation between the local behavior of the energy and some chosen operator \cite{filippi_optimal_2000,schautz_optimized_2004,scemama_simple_2006}. 
A set of operators ${\hat{A}_i}$ is a good set if only a few terms in Eqn~\ref{7_6} are non-zero, while a set with many small values in Eqn~\ref{7_6} is not an efficient descriptor of the wave function improvement. 

\section{QMC Methodology}

We first compute the single-particle Hartree-Fock (HF) orbitals for a molecular system. 
We obtain all orbitals using the GAMESS computational package \cite{schmidt_general_1993,gordon_chapter_2005}. 
Core electrons were replaced by the corresponding Burkatzki-Filippi-Dolg pseudopotential \cite{burkatzki_energy-consistent_2007} with triple-$\zeta$ basis sets. 

We perform variational Monte Carlo with the QWalk computational package \cite{wagner_qwalk:_2009}. We begin with a trial wave function of the Slater-Jastrow form:
\begin{equation}
\Psi = \text{exp}(U) \text{Det}[\phi_i(r_j)],
\label{sj_form}
\end{equation}
We use the linear method of Umrigar et. al. \cite{umrigar_energy_2005,umrigar_alleviation_2007, toulouse_optimization_2007} to optimize the Jastrow $U$. The form of the Jastrow correlation factor $U$ is a function of the electron and ionic coordinates:
\begin{equation}
U= \sum_{ijI} u(r_{iI}. r_{jI}, r_{ij}),
\end{equation} 
where $i$ and $j$ indices represent electronic coordinates and $I$ represents ionic coordinates. The functions $u$ are given by:
\begin{equation}
\begin{split}
u(r_{iI}, r_{jI}, r_{ij}) = \sum_k c_k^{ei} a_k (r_{iI})+ \\ \sum_m c_m^{ee} b_k (r_{ij}) + \sum_{klm} c_{klm}^{eei} (a_k(r_{iI})a_l(r_{jI}) + \\ a_k(r_{jI})a_l(r_{iI}))b_{k}(r_{ij}),
\end{split}
\end{equation}
where the $a_k$ and $b_k$ functions have the general form:
\begin{equation}
a_k(r) = \frac{1-z(r/r_{cut})}{1+ \beta z(r/r_{cut})},
\end{equation}
and $z(x)$ is a polynomial chosen to smoothly go to zero at $r=r_{cut}$ \cite{schmidt_correlated_1990}.  This form of the Jastrow factor explicitly incorporates three-body interactions between two electrons and an ion.

\section{Determinant selection}

The set of second-quantized two-body creation/destruction operators, 
\begin{equation}
\label{7_7}
\hat{A}_{ij,kl} \equiv c_{ \uparrow k}^{\dagger} c_{ \downarrow l}^{\dagger} c_{ \uparrow i} c_{ \downarrow j},
\end{equation}
offers one possible choice of linear operators $A_k$ in Eqn~\ref{7_6}. 
If $i,j$ are occupied orbitals and $k,l$ are unoccupied orbitals, then applying a $\hat{A}_{ij,kl}$ to a Slater determinant generates an additional excited-state determinant. 
The elements of the two-body reduced density matrix (2-RDM) are given by the expectation values of these two-body creation/destruction operators. 
We thus make the analogy with local energy to define a local density matrix element, given a wave function $|\Psi_{T}\rangle$:
\begin{equation}
\label{eq:10}
\rho_{ijkl} (\textbf{R}) = \frac{ \hat{A}_{ij,kl} \Psi_{T}(\textbf{R})}{ \Psi_{T}(\textbf{R}) }.
\end{equation}
Or, explicitly:
\begin{equation}
\begin{split}
\label{eq:11}
\rho_{ijkl} (\textbf{R}) = \sum_{a \neq b} \int \phi^{*}_{k}(r'_ a) \phi^{*}_{l}(r'_ b) \\ \times \phi_{i}(r_ a ) \phi_{j}(r_b ) \Psi^{*}_T (R''_{a b}) \Psi_T (R) dr'_{a} dr'_{b} .
\end{split}
\end{equation}
where $R=(r_1, r_2,...,r_N)$, $R''_{a b}=(r_1,r_2,...,r'_{a},...r'_{b},...,r_N)$ refers to the set of coordinates generated by changing the positions of two electrons, and we have omitted overall normalization. We evaluate this 2-body integral in a QMC calculation by sampling the coordinates $r'_{a}$ and $r'_{b}$ from the sum over orbitals $f(r) = \sum_i \phi_i^{2}(r)$ and the many-body electron coordinate $\textbf{R}$ from $\Psi^{2} (\textbf{R})$ \cite{wagner_types_2013}. With this, the expression given in Eqn~\ref{eq:11} can be rearranged to give:
\begin{equation}
\begin{split}
\label{2rdm_arrange}
\rho_{ijkl} (\textbf{R}) = \\ 
\frac{1}{N_i N_j N_k N_l} & \sum_{a \neq b} \left \langle \frac{ \frac{\Psi(\textbf{R}_{ab}'')}{\Psi(\textbf{R})} \phi^{*}_{k}(r'_ a) \phi^{*}_{l}(r'_ b) \phi_{i}(r_ a ) \phi_{j}(r_b )}{f(r'_a) f(r'_b)} \right \rangle _{f(r'_a),f(r'_b)} ,
\end{split}
\end{equation}
where the normalization factor is given by:
\begin{equation}
N_i = \sqrt{\left \langle \frac{\phi_i^2 (r'_a)}{f(r'_a)} \right \rangle_{f(r'_a)} }.
\end{equation}
The two particle operators in Eqn~\ref{7_7} are used to evaluate Eqn~\ref{7_6} and generate a list of important determinants missing from the initial wave function. Hence, we can select the determinants most important to the exact ground state without the need to first evaluate those determinants. The entire process of wave function generation is summarized as such:
.\begin{enumerate}
	\item Obtain single-particle orbitals from a HF calculation.
	\item Optimize single-determinant Slater-Jastrow:  
	\begin{equation}
	\langle \textbf{R} | \Psi \rangle  = e^{U(r,r')} \hbox{Det}[\phi_{i} (\textbf{r}_j)].
	\end{equation}
	\item Rank 2-RDM elements by covariance of $ \langle c_{ \uparrow k}^{\dagger} c_{ \downarrow l}^{\dagger} c_{ \uparrow i} c_{ \downarrow j} \rangle $ with $E_L$.
	\item Add corresponding determinant to the expansion:
	\begin{equation}
	  | \Psi_{\text{new}} \rangle  = | \Psi_{\text{old}} \rangle + a_{1} e^{U} [c_{ \uparrow k}^{\dagger} c_{ \downarrow l}^{\dagger} c_{ \uparrow i} c_{ \downarrow j}] | \hbox{Hartree-Fock} \rangle .
	\end{equation}
	\item Optimize coefficients $\{ c_i \}$ of $| \Psi_{\text{new}} \rangle $ using the linear method. 
\end{enumerate}
Iterating steps 3-5 of this process generates a determinantal expansion of arbitrary length up to the full size of the active space. 

\subsection{H$_2$ molecule}

\begin{figure}[t!]
	\includegraphics{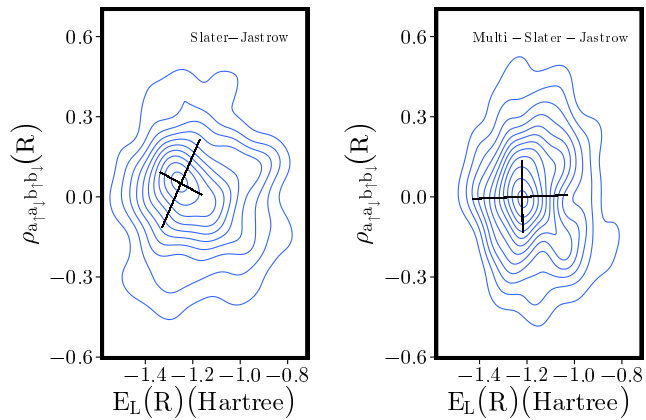}
  \caption{Amplitude of 2-body $b \rightarrow a$ bonding-to-antibonding excitation $\rho_{a_{\uparrow} a_{\downarrow} b_{\uparrow} b_{\downarrow}}(\textbf{R})$ versus local energy $E_L(\textbf{R})$ for two different trial wave functions, with corresponding principal components of the distribution indicated. The Slater-Jastrow wave function used to generate the left panel did not include the CSF corresponding to this 2-body excitation, while the wave function used to generate the right panel does. The principal components rotate upon the addition of this CSF.}
	\label{h2_shift}
\end{figure} 

For the case of H$_2$, we restrict our active Hilbert space to the set of bonding/antibonding $\sigma$-symmetry orbitals. Fig. \ref{h2_shift} shows the contours of the sampled amplitude $\rho_{a_{\uparrow} a_{\downarrow} b_{\uparrow} b_{\downarrow}}(\textbf{R})$ of the local operator associated with a 2-body $b\rightarrow a$  bonding-to-antibonding excitation in an isolated hydrogen dimer versus the sampled local energy $E_{L} (\textbf{R})$ for each of two trial states:
\begin{equation}
\begin{split}
\Psi_{SJ} & = e^U \phi_{b \uparrow}(r_1) \phi_{b \downarrow} (r_2)\\
\Psi_{MSJ} & = e^U (c_1 \phi_{b \uparrow}(r_1) \phi_{b \downarrow} + c_2 \phi_{a \uparrow}(r_1) \phi_{a \downarrow} ) , 
\end{split}
\end{equation}
where $\Psi_{SJ}$ and $\Psi_{MSJ}$ are the Slater-Jastrow and multi-Slater Jastrow wave functions containing the bonding $\phi_b$ and antibonding $\phi_a$ single-particle orbitals respectively.

The line segments on each panel in Fig \ref{h2_shift} indicate the principal components of the resulting distribution. These components are given by the eigenvectors of the covariance matrix of the local energy distribution taken with respect to the local operator $\rho_{a_{\uparrow} a_{\downarrow} b_{\uparrow} b_{\downarrow}}(\textbf{R})$:
$$
\quad
\begin{pmatrix}
\sigma_{\rho , \rho} & \sigma_{\rho,E_L} \\
\sigma_{E_L,\rho} & \sigma_{E_L , E_L}
\end{pmatrix}
$$
in this two-dimensional representation. 
The rotation of the principle components relative to the axes in the left panel of Fig. \ref{h2_shift} shows that the covariance matrix contains nonvanishing off-diagonal elements. 
It follows that $\rho_{a_{\uparrow} a_{\downarrow} b_{\uparrow} b_{\downarrow}}(\textbf{R})$ and $E_L(\mathbf{R})$ are correlated for this single-determinant trial state. 
After the addition of the associated $b \rightarrow a$ determinant to the wave function in the right panel of Fig. \ref{h2_shift}, the principle components rotate to align with the axes, indicating that the covariance matrix has become diagonal. This implies that the covariance between the local energy $E_L(\mathbf{R})$ and local operator $\rho_{a_{\uparrow} a_{\downarrow} b_{\uparrow} b_{\downarrow}}(\textbf{R})$ has vanished, and the two variables now have zero covariance. That is, a key element absent from the initial trial state has been identified and added based on the covariance, pushing the wave function closer to the exact ground state. 

\subsection{Dimer Molecules}

As a further proof of concept, we apply the covariance method to select determinants for a set of stretched molecules: H$_2$ (0.88 {\AA} bond length), N$_2$ (1.7 {\AA} bond length), O$_2$ (1.6 {\AA} bond length), and F$_2$ (1.5 {\AA} bond length). 
By stretching the molecules, the electron correlations are enhanced, increasing the strength of the covariance signal.
We obtain single-particle orbitals for each system from a restricted open-shell Hartree-Fock (ROHF) calculation using GAMESS. This method doubly-fills molecular orbitals (MOs) to the greatest extent possible, and places remaining unpaired electrons into singly-filled MOs. We limit our active space to a set of bonding and antibonding MOs with cylindrical symmetry and either $\sigma$- or $\pi$-symmetry. Other states exist within the full orbital space, but their inclusion yields only small improvement to the final wave function and system energy. Because different methods of determinant selection produce significantly different rates of energy convergence \cite{iii_influence_2015}, the covariance-based method we have described can yield interesting results even at the level of a multi-Slater-Jastrow \textit{ansatz}. Our chief objective in this section is to show that the covariance technique can select the most significant determinants for a particular molecule before performing a variational optimization of the wave function.

\begin{figure}
	\includegraphics{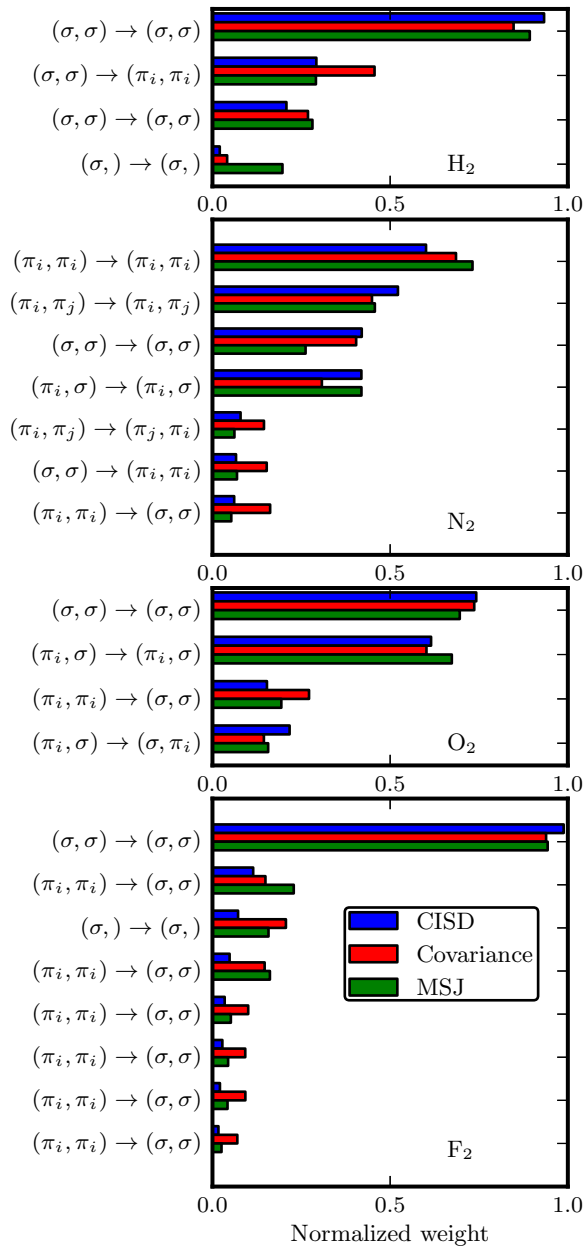}
  \caption{Comparison of normalized signal strength for different estimators of relative CSF importance for stretched dimers of H$_2$, N$_2$, O$_2$, and F$_2$ respectively. The CSFs are arranged such that the optimized final CSF weight declines monotonically from top to bottom. Each indicated excitation is a 1- or 2-particle excitation that includes both itself and any symmetry-related partners. For example, $(\pi_i,\pi_i) \rightarrow (\pi_i,\pi_i)$ is a 2-particle excitation that excites a bonding $\pi$-orbital electron to an antibonding $\pi^*$ orbital of the same angular momentum ($x$ or $y$) in each spin channel. On the other hand, $(\pi_i,\pi_j)\rightarrow(\pi_j,\pi_i)$ involves a two-body exchange.}
	\label{estimators_all}
\end{figure} 

\begin{figure}
	\includegraphics{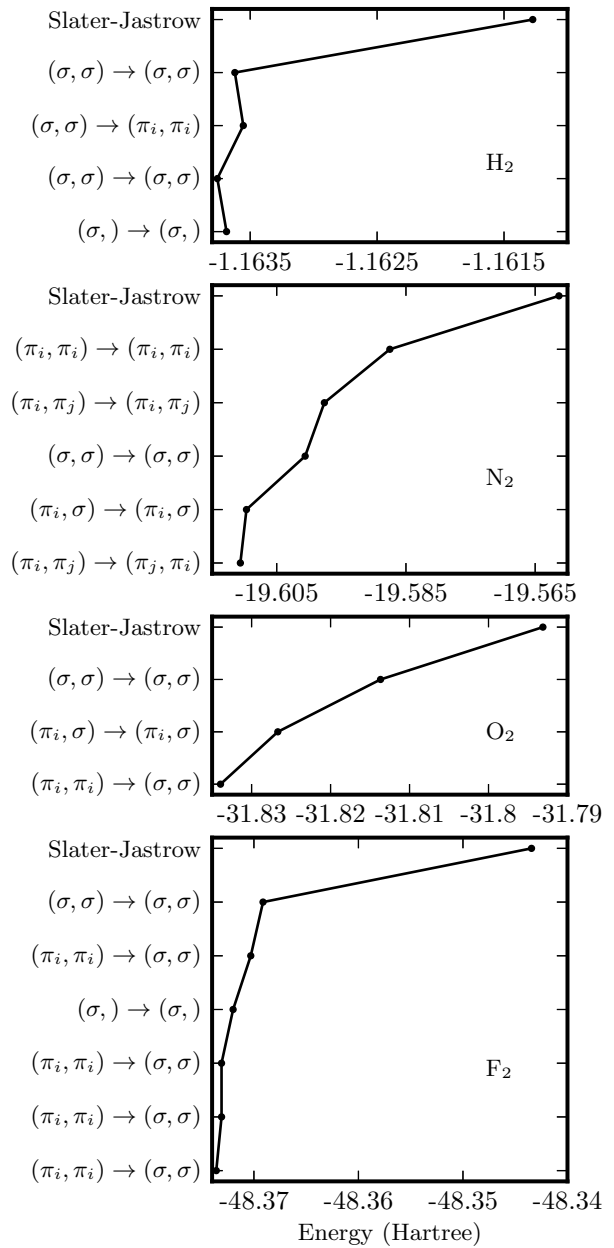}
  \caption{Added spin-up/spin-down CSF excitations vs. associated variational Monte Carlo energy in a multi-Slater-Jastrow wave function for the CSF ordering suggested by conventional CISD for each considered model system.}
	\label{energy_comp}
\end{figure} 

\begin{figure}
	\includegraphics{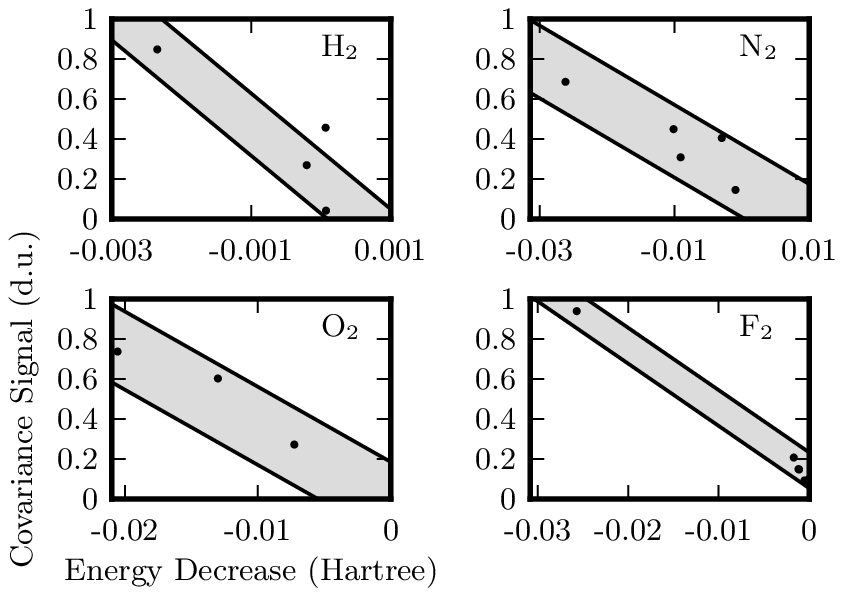}
	\caption{The normalized covariance signal of CSFs versus the decrease in energy obtained from adding a CSF to the trial state. Significant negative correlations exist between the two values. The shading is provided as a visual guide.}
	\label{energy_deriv}
\end{figure} 

%Using an optimized Slater-Jastrow wave function of the form seen in Eqn~\ref{sj_form}, we compute the local 2-body excitation operator $\rho_{ijkl} (\textbf{R})$, local energy $E_{L} (\mathbf{R})$, and associated covariance. We then rank-order the 1- and 2-body orbital excitations based on our existing scheme. 

We consider only 2-particle excitations featuring 1 particle in each spin channel. We compare these results to those obtained with the usual configuration interaction method with singles and doubles excitation (CISD). This is natural for molecules such as N$_2$ with a ground state singlet spin configuration, though it can lead to the exclusion of significant excitations in molecules like O$_2$ which contain unpaired electrons. Fig. \ref{estimators_all} compares the normalized weight of each CSF in conventional CISD, the optimized weight of each CSF in a multi-Slater-Jastrow ansatz, and the local energy covariance for each relevant CSF in each material respectively. We see that the determinant orderings predicted by both traditional CISD and our method based on local energy covariance are equivalent for each system across the dominant particle excitations. This indicates that the path to the ground state through Hilbert space obtained by successively applying the projection operator is approximately equivalent to that produced by the usual CI procedure in this case. 

From Eqn~\ref{7_6}, we see that the covariance signal in a 2-RDM element should fall identically to zero once the corresponding excitation has been added to the trial state. In practice, we observe that the signal in an added excitation falls significantly once it has been added to the trial wave function, but it does not vanish entirely. 
This is a consequence of the Jastrow factor in the trial state, which spoils the orthogonality of determinants and prevents the covariance in each 2-RDM element from fully vanishing. That is, the overlap matrix $S_{ik}$ is non-diagonal when a Jastrow factor multiplies the determinant expansion. Practically speaking, this slight non-orthogonality did not seem to affect the performance of the technique. Assuming that $S_{ik}$ is diagonal allows the method to scale linearly with the number of excitation operators, so we use this approximation. 

%As an additional benefit, the assumption of approximate orthogonality grants a linear scaling in the number of excitations operators $N_A$. That is, the number of determinants in a wave function expansion self-evidently scales as $\mathcal{O}(N_A)$. Dropping the orthogonality condition disrupts this linearity. . 

%As an additional benefit, the assumption of approximate orthogonality grants a linear scaling to our method with respect to the size of the wave function expansion. Alternatively, one could check the covariance of each 2-particle excitation individually, add the determinant corresponding to the largest covariance to the expansion, and recompute the 2-RDM of the new wave function before iterating the procedure. This modified procedure would also permit the generation of multi-Slater-Jastrow wave functions of arbitrary length. However, this modification disrupts the linear scaling of the technique and adds an additional step to our algorithm that appears to provide little practical benefit in terms of the excitations identified as significant. 

We also find the rate of energy convergence for the predicted CSF ordering in each model molecular system. Fig. \ref{energy_comp} shows the variational Monte Carlo energy of an optimized multi-Slater-Jastrow wave function as a function of the CSFs included in the trial state. The CSFs are ordered here according to the weight given by a conventional CISD calculation. We see that the energy converges rapidly with respect to the number of CSFs included in the wave function. This explicitly illustrates that the CISD method and our covariance-based technique can drive the initial trial state asymptotically close to the exact ground state. 

Finally, we also assess the degree to which the covariance in a 2-RDM element predicts the energy gain obtained from adding the associated determinant to the trial state. 
Fig. \ref{energy_deriv} compares the decrease in total system energy obtained from each additional CSF with the corresponding covariance signal. 
We observe that the energy gain and the covariance signal are negatively correlated with one another. 
This correlation indicates that the covariance in a 2-RDM element can be used as a proxy for estimating the energy change from adding a determinant to the trial state. 

As a method of determinant selection for these systems, this technique is less efficient than using CI to determine the weights, and the results are similar.
We therefore would not recommend this technique as a selection method for small molecules.
However, the point of this section is that the energy fluctuations can be data mined to find the correct directions in Hilbert space to improve trial wave functions.
In the case of stretched dimers, it is well-known that the most important improvement over Slater-Jastrow consists of multiple determinants, and the energy fluctuation technique selects the correct ones.

\section{Comparing real and orbital spaces: TiO molecule}

We now proceed to use the technique to selectively improve wave function parameterizations in a more challenging case.
As an example of a system where we do not know \textit{a priori} the most important degrees of freedom, we consider a transition metal molecule, TiO. 
The dynamic correlation present in transition metal systems is larger than in s-p systems like the dimers considered above, so the Jastrow factor could be expected to play a larger role\cite{visscher_electronic_1993,roos_new_2005}. 

\begin{figure}
  \includegraphics{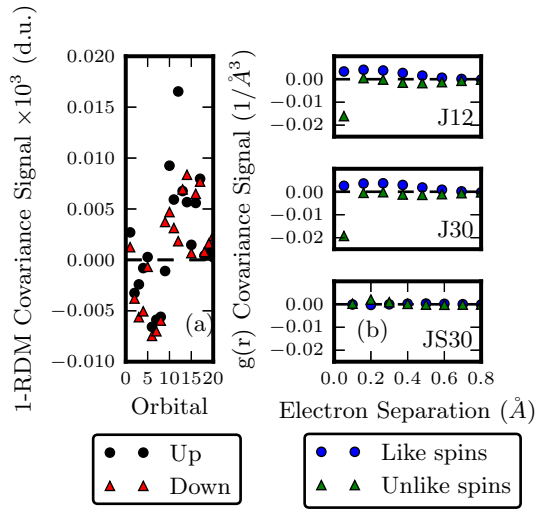}
  \caption{(a.) The covariance of the 1-RDM with the local energy $E_L$ for TiO in the J12 wave function. (b.) The covariance of the pair distribution $g(\textbf{r})$ with the local energy $E_L$ for the J12, J30, and JS30 wave functions in both spin channels (right). J12: Slater-Jastrow state with 12 parameters per atom in the 3-body part of the Jastrow factor; J30: J12, but with 30 3-body terms instead of 12; JS30: J30, but with a spin-dependent 2-body portion of the Jastrow factor.}
	\label{fig:gr_and_1rdm}
\end{figure} 

\begin{figure}
  \includegraphics{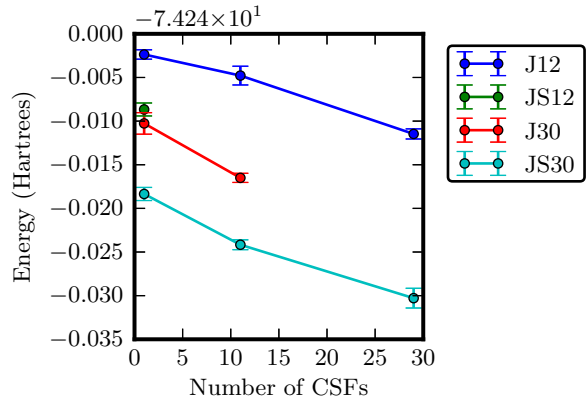}
  \caption{Total TiO VMC energy vs. the number of included CSFs using different Jastrow factors. Note that the decline in energy is quite modest with respect to the number of included CSFs, but falls dramatically when spin-dependence is incorporated in the 2-body portion of the Jastrow factor.}
	\label{fig:tio_energy}
\end{figure} 

In this case, the 1-RDM can be used to select the orbital space by calculating the covariance of the diagonal elements of the 1-RDM with the local energy. 
In the case of the stretched N$_2$ dimer, the elements corresponding to the bonding and antibonding orbitals have a covariance with the local energy of approximately 0.001, while other orbitals have much smaller signals.

In Fig~\ref{fig:gr_and_1rdm}, the covariances of the 1-RDM and the real-space electron-electron correlation function $g(\textbf{r})$  distance are shown. 
The covariance signal for the 1-RDM is very small, and in fact the selection of determinants was very difficult in this case, although we do obtain larger signals for the $p$ and $d$ states as one would expect.
On the other hand, for our starting wave function, labeled J12, with 12 three-body parameters per atom, there is a large spin-dependent covariance with $g(\textbf{r})$. 
Increasing the number of parameters in the Jastrow factor without making it spin-dependent (J30, with 30 3-body parameters per atom) does not resolve this covariance, but adding in four spin dependent parameters (JS30) immediately reduces the covariance signal to nearly zero across all separation distances.

Since the determinant selection of TiO via energy covariance was not efficient, we used a CI calculation with sextuple excitations into 8 virtual states to select CSFs, then formed a set of multi-Slater Jastrow wave functions.
If the covariance analysis was correct, then we would expect the spin-dependent terms in the Jastrow to be most effective in lowering the energy, followed by either the extra three-body terms or multiple determinants.
As can be seen in Fig~\ref{fig:tio_energy}, this supposition is correct: with only four parameters, the spin-dependent terms lower the energy by nearly 10 mHartree, while 30 determinants or a similar number of 3-body parameters are necessary to achieve that decrease in energy.

This example illustrates some the strengths and weaknesses of this covariance-based selection.
If the set $\{ A_i \}$ is selected in a basis that does not describe the needed improvement efficiently, in this case the determinant basis, then it is not the best tool.
On the other hand, if several different basis sets are used, then the best basis can be used to improve the wave function. 
In this case, we learned that a spin-dependent Jastrow factor can improve the energy significantly for magnetic molecules, while the determinant basis is not an efficient way to improve the wave function for this molecule.
The cost for performing these calculations was about a factor of two larger than a variational Monte Carlo calculation and much smaller than the energy optimization technique. 

\section{Conclusion}
We have presented an outline of a technique to select, not just terms in a many-body ansatz, but which type of ansatz with which to proceed.
For example, the selection method can quickly determined whether a determinant-type basis is appropriate by evaluating the 1-RDM covariance with the local energy.
Similarly, if an explicitly correlated approach such as a Jastrow is more appropriate, then the covariance of the local energy with the electron-electron distance $g(r)$ is large.
The computational cost of this assessment is quite low: $g(r)$ is essentially zero cost over a VMC energy evaluation, and the 1-RDM is approximately a factor of two additional, regardless of system size. This is much less expensive than attempting energy minimization on multiple ansatz.

As proof of concept, we demonstrated that the selection technique both selects the correct directions in Hilbert within a defined ansatz space, and also can select between alternate viewpoints of the electron correlation problem.
We demonstrated the former by selecting determinants for stretched dimer molecules, and the latter by differentiating between short range 'dynamic' correlation best described by a Jastrow factor and long range 'static' correlation best described by multiple determinants in the transition metal oxygen system TiO. Using standard wave functions for this problem, the dynamic correlation in TiO is more important. This work forms the base for an algorithm in which the local energy can be analyzed directly in the many-body space using feature extraction techniques to describe the most efficient basis in which to improve many-body wave functions.

This material is based upon work supported by NSF DMR-1206242 (L.K.W.) and the National Science Foundation Graduate Research Fellowship Program under Grant Number DGE-1144245 (K.T.W.). 
We also acknowledge computer resources from the Campus Cluster program at Illinois. 
Useful conversations with David Ceperley are gratefully acknowledged.

\bibliography{sources5}

\end{document}